# Magnetic and structural study of Cu-doped TiO$_2$ thin films

C.E. Rodríguez Torres[a]*, F. Golmar[b], A.F. Cabrera[a], L.A. Errico[a], A.M. Mudarra Navarro[a], M. Rentería[a], F.H. Sánchez[a], and S. Duhalde[b]

[a]*Dpto de Física-IFLP, Fac. Cs. Exactas, Universidad Nacional de La Plata-CONICET, CC 67, 1900 La Plata, Argentina.*

[b]*Laboratorio de Ablación Láser, Facultad de Ingeniería, Universidad de Buenos Aires, Paseo Colón 850, 1063 Buenos Aires, Argentina.*

\* Corresponding author. Tel.: +54-221-4246062; fax: +54-221-4252006.
*E-mail address*: torres@fisica.unlp.edu.ar.

Abstract

Transparent pure and Cu-doped (2.5, 5 and 10 at. %) anatase TiO$_2$ thin films were grown by pulsed laser deposition technique on LaAlO$_3$ substrates. The samples were structurally characterized by X-ray absorption spectroscopy and X-ray diffraction. The magnetic properties were measured using a SQUID. All films have a FM-like behaviour. In the case of the Cu-doped samples, the magnetic cycles are almost independent of the Cu concentration. Cu atoms are forming CuO and/or substituting Ti in TiO$_2$. The thermal treatment in air promotes the CuO segregation. Since CuO is antiferromagnetic, the magnetic signals present in the films could be assigned to this Cu substitutionally replacing cations in TiO$_2$.

Introduction

A variety of doped-semiconducting materials, called diluted magnetic semiconductors (DMS) combine two properties: semiconductor nature and ferromagnetic behaviour. DMS materials which retain their magnetism at and above RT are critically important in the development of spintronics as spin injectors for semiconductor heterostructures that can operate without cryogenic cooling [1]. Group IV, III-V and II-VI DMS materials typically exhibit Curie temperatures ($T_C$) well below room temperature (RT) [2]. But, in the last years, it has been shown that Co-doped $TiO_2$ films (with rutile and anatase structures) are ferromagnetic well above RT ($T_C > 400$ K) [3]. After these reports, many research groups have focused their work on doping $TiO_2$ with transition metals. However, the origin of magnetic moment, the coupling mechanism, and the precise state of the impurities remain unknown.

Recently, we reported unexpected and significant RT ferromagnetism in Cu-doped TiO2 films [4], equivalent to what would be expected if, on the average, each Cu atom bears a magnetic moment of about 1.5 $\mu_B$. This experimental result indicates that ions capable to develop ferro or ferrimagnetic order are not essential to obtain this effect and also that this is not due to impurity clustering. Ab initio calculations on bulk Cu-doped $TiO_2$, with and without oxygen vacancies, predict a magnetic moment of 1.0 $\mu_B$ for a supercell (a´=b´=4.58451 Å, c´=2c=5.90662 Å, see ref. 4) containing one Cu impurity and a neighbour oxygen vacancy, but no magnetic moment is shown if the oxygen vacancy is absent. The calculations also predict lower vacancy formation energies due to the presence of Cu impurities.

In this work we present a study of the magnetic and structural properties of pure and Cu-doped $TiO_2$ thin films deposited on $LaAlO_3$ by pulsed laser deposition (PLD), in order to understand the role of the dopant in the origin and significance of the observed ferromagnetism.

Experimental

Thin films of pure TiO$_2$ and 2.5, 5.0 and 10 at. % Cu-doped TiO$_2$ were deposited on LaAlO$_3$ (001) substrate by PLD using a Nd:YAG laser operating at 266 nm. The substrate temperature, laser energy density, oxygen pressure, and pulse repetition rate were 800 °C, 2 J/cm$^2$, 20 Pa, and 10 Hz, respectively. The samples were thermally treated at 800 °C during 30 min in air in order to determine the role played by the oxygen vacancies. X-ray absorption spectroscopy (XAS), X-ray diffraction (XRD) and SQUID measurements were performed in the as-deposited and the thermal treated samples. In all cases, magnetisation vs. applied field measurements were performed with the external field applied parallel to the plane of the film. XAS (XANES: x-ray absorption near edge spectroscopy and EXAFS: extended x-ray absorption fine structure) measurements were taken at room temperature in fluorescence mode at the Cu K-edge for all Cu-doped samples, using the Si (111) monochromator at the XAS beamline of LNLS (Campinas, Brazil).

Results and discussion

All the films resulted transparent to the visible light and strongly textured. X-ray diffraction (XRD) studies (not presented here) showed only the (001) reflection of the anatase structure for pure and Cu-doped TiO$_2$ films in the as-deposited state and after the thermal treatment.

Fig. 1 (top) shows the magnetic hysteresis loops of the pure and as-deposited Cu-doped TiO$_2$ films. All films have a FM-like behaviour. As can be seen, no appreciable differences can be distinguished in the magnetic cycles corresponding to the

2.5%, 5% and 10 at.% Cu-doped samples. The estimated saturation magnetizations (calculated supposing a film of 150 nm thick) are 35 ± 3, 37 ± 3 and 43 ± 3 emu/cm$^3$, respectively. In the case of the pure film, the observed ferromagnetic-like behaviour (with an estimated saturation magnetization of 19 ± 3 emu/cm$^3$) is intriguing, but not unexpected, since similar results were reported by N. Hong et al [5]. However, in the present case, the pure film ferromagnetism almost disappeared after keeping the sample in air during two months.

After the thermal treatment of the samples, the saturation magnetization of the doped films decreases as shown in Fig 1 (bottom) and reaches a saturation value similar to the one observed for the pure film (measured immediately after preparation). Contrary to what was observed for the pure film, the ferromagnetic-like behaviour of Cu doped films, as-deposited and thermally treated, remains in time. M(T) curves taken under 1T applied field (not shown here) are almost constant in the temperature range 10-300 K, suggesting a $T_C$ considerably higher than 300 K.

Fig. 2 shows the Cu K-edge EXAFS spectrum (top) and their corresponding Fourier transform (bottom) of the 10 at.% Cu-doped TiO$_2$ film as-deposited and after the previously mentioned thermal treatment. No appreciable differences were observed between this spectrum and those corresponding to the other Cu-concentrations studied here, therefore, we will refer our discussion to the 10 at.% Cu-doped TiO$_2$ film. We also present in Fig. 2 the EXAFS spectrum obtained for copper monoxide (CuO). As can be seen, there is a high degree of similarity between the EXAFS oscillations of Cu-doped TiO$_2$ film and those of CuO. This similarity is also observed in the Fourier transform (FT). This is an evidence that Cu atoms are in a CuO-like local structure, suggesting that in these films most of the Cu atoms are forming CuO precursor. However, Fig. 2 also shows that the anatase FT (taken from a Ti K-edge measurement) is also quite similar to that of CuO, which makes more difficult to distinguish between Cu in CuO and Cu (substitutional) in anatase $TiO_2$.

The Cu-K edge XANES spectra shown in Fig. 3 (top) also indicates that the chemical environment around the Cu atoms in the as-deposited films is similar to that of CuO, although a small shift toward higher energies is observed. The edge position is coincident with that of $CuTiO_3$ [6], suggesting that some Ti atoms are in the vicinity of Cu. In order to make more visible the features present in the steeply rising edge, we subtracted an arctan function from the original data sets. The resulting patterns are shown in Figure 3 (bottom). Based in the analysis of Lytle et al. [6] we associate the peak at 35 eV with single scattering (SS) from oxygen atoms located at approximately 1.96 Å (in average) from Cu. This Cu-O distance is in agreement with the Cu-O distances in CuO and also with the Cu-O distances predicted by *ab initio* calculations for Cu substitutionally located in rutile and anatase $TiO_2$ [4] and [7]. At this point it is interesting to note the existence of a small peak located at 15 eV in the as- cast film which is no longer present in the thermally treated one. This peak can be associated with single scattering from atoms located at 3.05 Å. This distance is very similar to the Cu-Ti nearest neighbour distance for Cu replacing Ti in rutile and anatase $TiO_2$. This fact and the larger contribution at 2.8 Å in the FT suggest that some amount of Cu could be substitutionally located in $TiO_2$ and some of it segregated in CuO after the thermal treatment. This idea is supported by the fact that, after the thermal treatment, the edge shifts towards lower energies coinciding with the Cu K-edge for CuO. Since the Cu oxides are antiferromagnetic, the magnetic signals present in the films could be assigned to this Cu substitutionally replacing cations in $TiO_2$, in agreement with our *ab initio* calculations [7]. In the same way, the decrease of the magnetization after the thermal treatment can be associated with CuO segregation.

Conclusions

All films have a FM-like behaviour and no appreciable differences can be distinguished between the magnetic cycles corresponding to samples doped with different concentrations of Cu (in the range 2.5 – 10 at.% of Cu). Cu atoms are forming CuO precursors and/or substituting Ti in $TiO_2$. The thermal treatment promotes the CuO segregation. Since CuO is antiferromagnetic, the magnetic signals present in the films could be assigned to Cu substitutionally replacing cations in $TiO_2$. In this sense the decrease of magnetic signal after the thermal treatment can be associated with CuO segregation.

Acknowledgments

Research grants PIP 5006 and PIP 6032 from Consejo Nacional de Investigaciones Científicas y Técnicas (CONICET, Argentina), Fundación Antorchas, Red Nacional de Magnetismo (RN3M) and ANPCyT are gratefully acknowledged. This work was dedicated to the memory of Dra. Stella Duhalde who passed away recently.

Figure Captions

Figure 1: (top) Magnetization curves of pure and Cu-doped $TiO_2$ films measured at RT with the magnetic field applied parallel to the film surface for the different studied dopant concentrations. (bottom) Curves comparing the magnetization from Cu 2.5 % film as-deposited (AD), the same sample after a thermal treatment (TT) in air, and a pure $TiO_2$ film. In all cases the substrate signal was subtracted.

Figure 2: EXAFS spectrum (top) and their corresponding Fourier transform (bottom) of a 10 at. % Cu-doped $TiO_2$ film, as-deposited (AD), and after a thermal treatment in air (TT). The results for CuO and anatase $TiO_2$ are also shown for comparison.

Figure 3: (top) XANES spectra of 10 at. % Cu-doped $TiO_2$ as-deposited (AD); (bottom) Results of subtracting an arctan function from the original dataset (see text). The results obtained for CuO and $CuTiO_3$ are also shown for comparison.

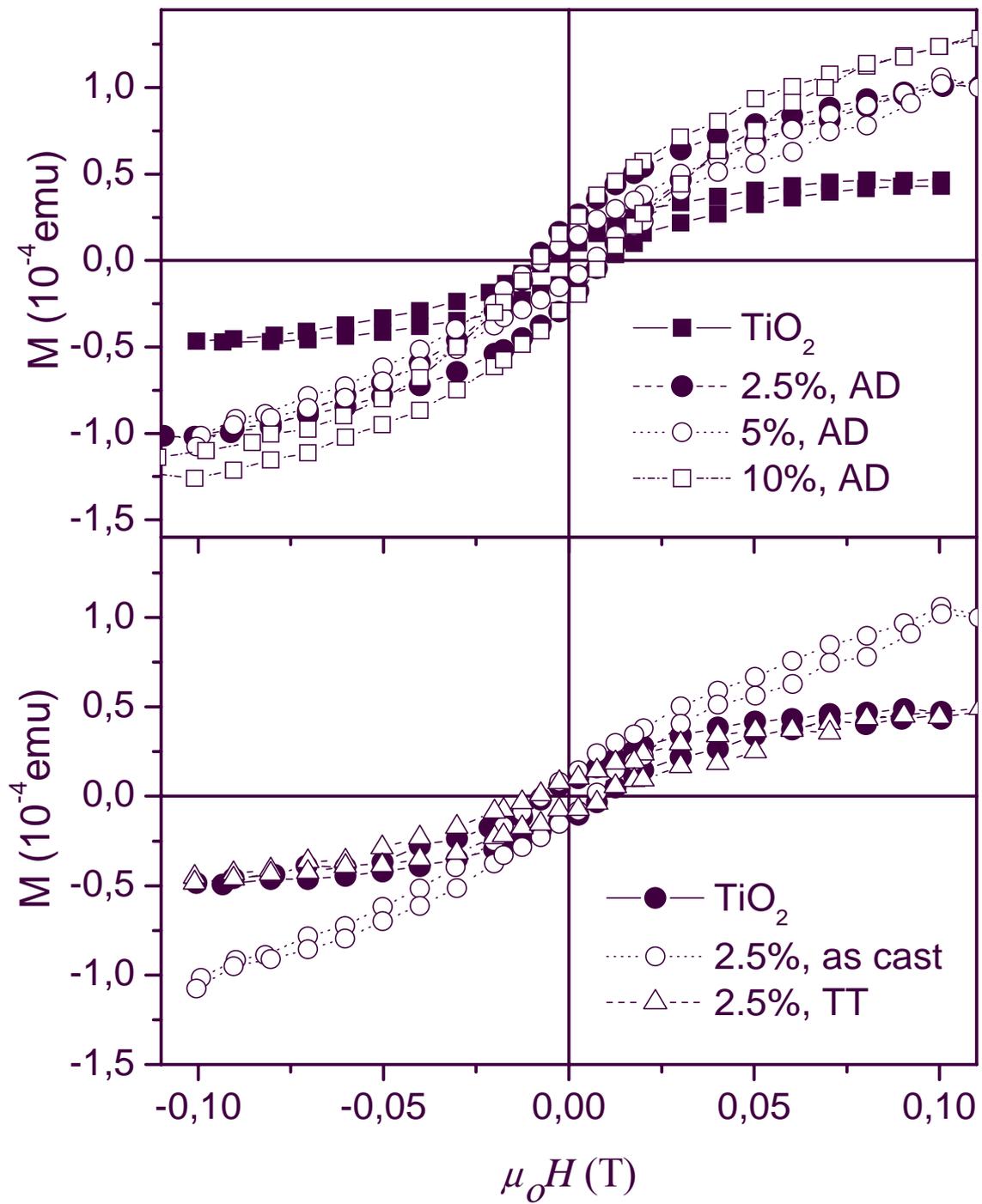

Figure 1

**Figure 2**

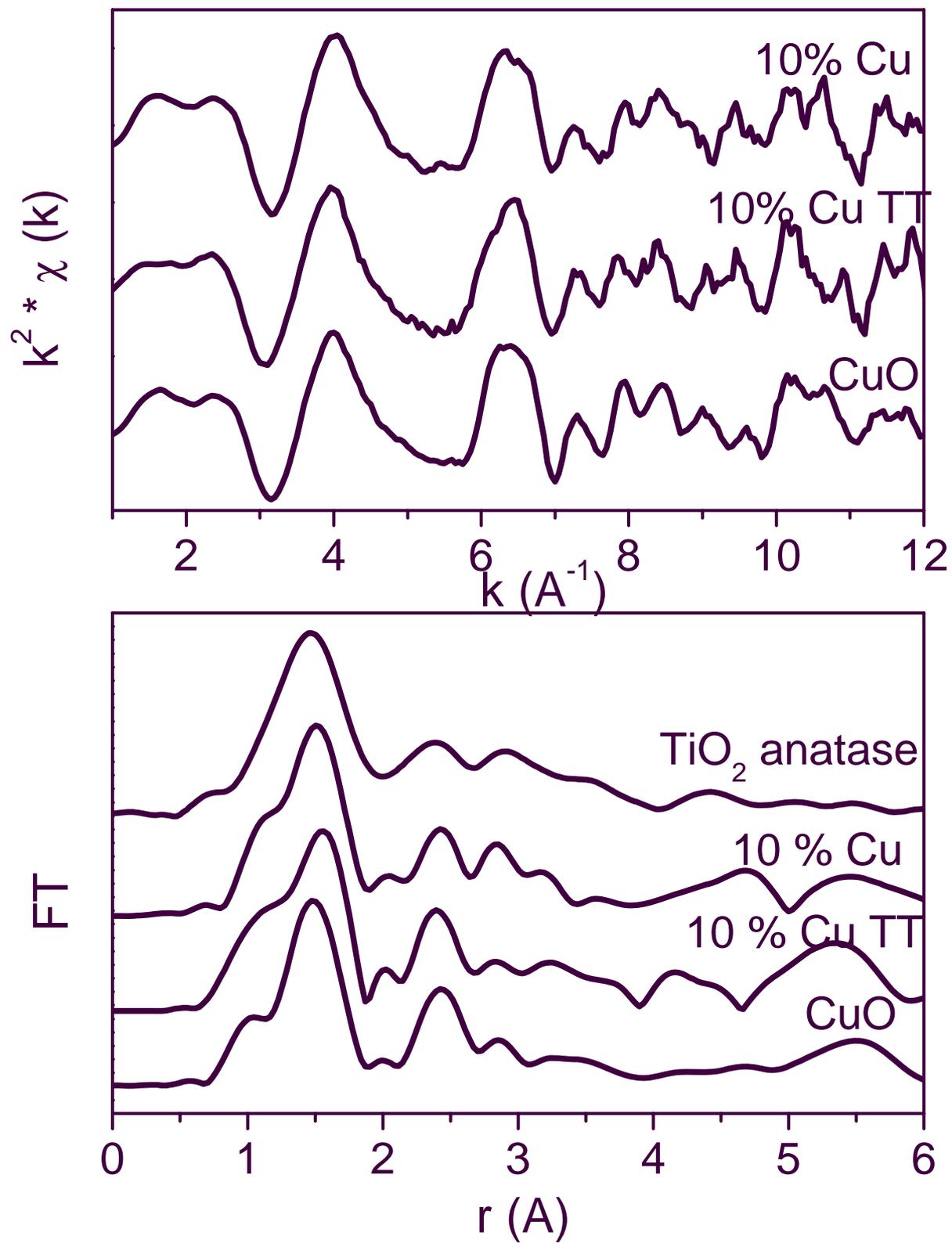

Figure 3

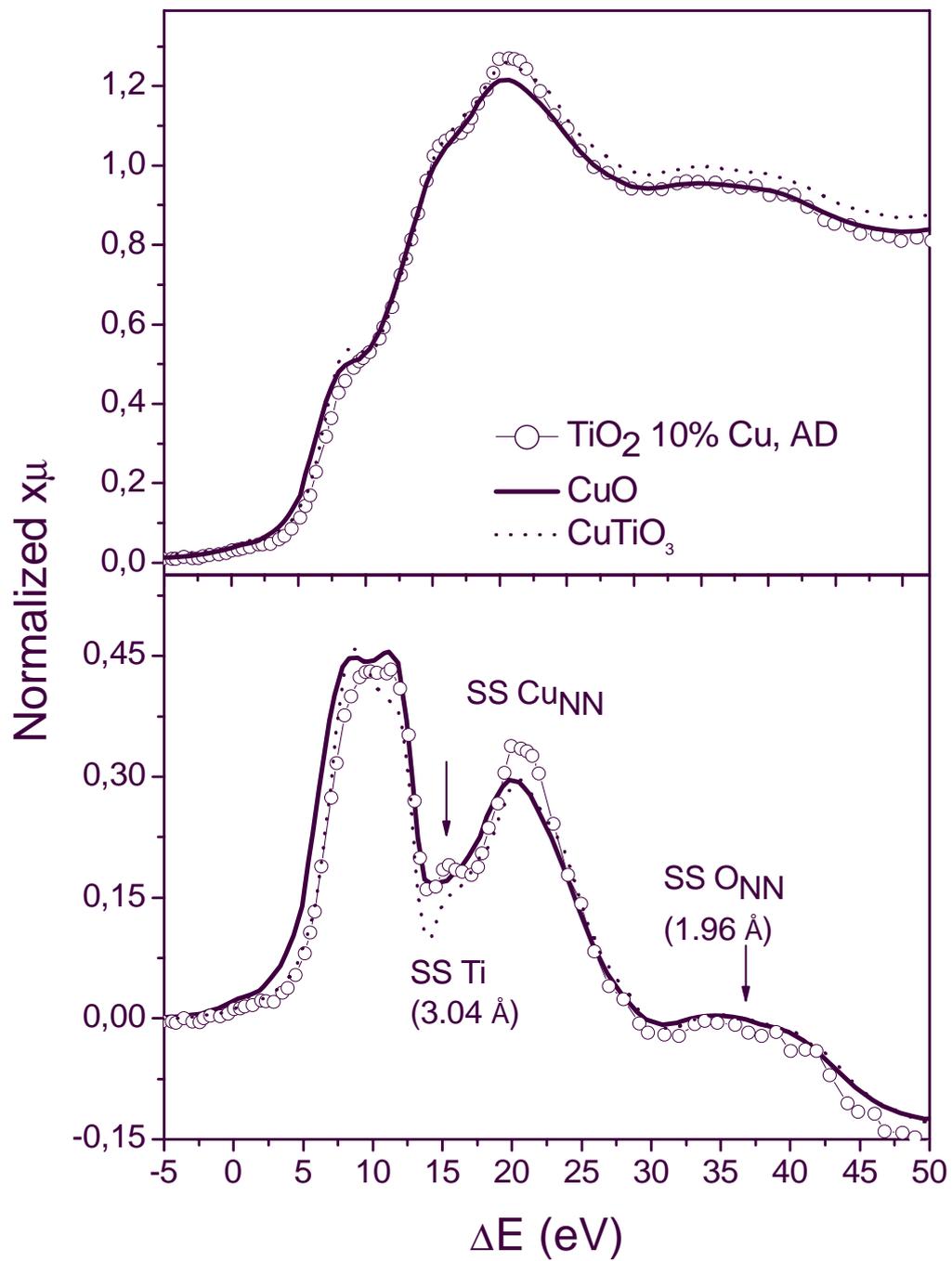